\documentclass[sigconf]{acmart}
\usepackage{siunitx}

\usepackage{multirow}
\usepackage{makecell}
\usepackage{wrapfig}

\usepackage{rotating,tabularx,adjustbox}

\AtBeginDocument{%
  \providecommand\BibTeX{{%
    \normalfont B\kern-0.5em{\scshape i\kern-0.25em b}\kern-0.8em\TeX}}}

\copyrightyear{2024}
\acmYear{2024}
\setcopyright{acmlicensed}\acmConference[SIGCSE 2024]{Proceedings of the 55th ACM Technical Symposium on Computer Science Education V. 1}{March 20--23, 2024}{Portland, OR, USA}
\acmBooktitle{Proceedings of the 55th ACM Technical Symposium on Computer Science Education V. 1 (SIGCSE 2024), March 20--23, 2024, Portland, OR, USA}
\acmPrice{15.00}
\acmDOI{10.1145/3626252.3630873}
\acmISBN{979-8-4007-0423-9/24/03}

\begin{document}

\title{Implementation of Split Deadlines in a Large CS1 Course}


\author{Hongxuan Chen}
\orcid{0000-0002-1606-0484}
\affiliation{%
  \institution{University of Illinois at Urbana-Champaign}
  \streetaddress{201 N. Goodwin Ave}
  \city{Urbana}
  \state{Illinois}
  \country{USA}
  \postcode{61801}
}
\email{hc10@illinois.edu}

\author{Ang Li}
\affiliation{%
  \institution{University of Illinois at Urbana-Champaign}
  \streetaddress{201 N. Goodwin Ave}
  \city{Urbana}
  \state{Illinois}
  \country{USA}
  \postcode{61801}
}
\email{angl2@illinois.edu}

\author{Geoffrey Challen}
\orcid{0000-0002-7249-7867}
\affiliation{%
  \institution{University of Illinois at Urbana-Champaign}
  \streetaddress{201 N. Goodwin Ave}
  \city{Urbana}
  \state{Illinois}
  \country{USA}
  \postcode{61801}
}
\email{challen@illinois.edu}

\author{Kathryn Cunningham}
\orcid{0000-0002-9702-2796}
\affiliation{%
  \institution{University of Illinois at Urbana-Champaign}
  \streetaddress{201 N. Goodwin Ave}
  \city{Urbana}
  \state{Illinois}
  \country{USA}
  \postcode{61801}
}
\email{katcun@illinois.edu}





\begin{abstract}

Office hour utilization in computer science courses can spike near deadlines, producing long wait times, frustrated students, and overworked staff.
To address this problem, a large CS1 course implemented a \textit{split deadlines} policy.
Students were randomly divided into two groups with staggered release and due dates.
Each group had the same amount of time to complete assignments, but the number of students with each due date was reduced by half.

Our study evaluates the effectiveness of this policy.
We measure office hour utilization and staff efficiency near deadlines, examine the policy's impact on student performance, and investigate student perception of the policy's fairness and effectiveness.
Overall we found that the split deadline policy increased office hour efficiency, resulted in no significant difference in performance between groups, and was considered fair and effective by most students.
Our experience report includes reflections and student feedback indicating how to implement and further improve similar policies.

%
%

\end{abstract}



\maketitle

\newcommand{\OH}{office hour}
\newcommand{\OHS}{office hours}
\newcommand{\Intro}{CS~124}

\section{Introduction}

Office hours---set times when students can receive support from course staff---play an important role in supporting students in computer science courses~\cite{wirtz2018resource,doebling2021patterns,smith2017office}.
By allowing students to receive individual or group tutoring from course staff, \OHS{} can both support learning and nurture a sense of course community.

However, rising enrollments in CS courses are complicating the challenge of providing \OHS{}~\cite{roberts2018rising,camp2017growth}.
Large courses may struggle to find enough space or staff to run \OHS{} effectively---resulting in long waits, overcrowding, and a negative environment for staff and students.
Utilization of \OHS{} is also frequently unevenly distributed.
Most students tend to work on assignments near deadlines, and \OHS{} that were sparsely attended all week can become packed on the due date~\cite{smith2017office}, with the impending deadline worsening an already stressful atmosphere.

Packed rooms and long lines result from an imbalance between staff availability (supply) and student requests (demand).
Many existing efforts to improve \OHS{} focus on increasing or improving \textit{supply}---including assigning more staff to work around deadlines, and the development of tools to increase staff efficiency~\cite{smith2017my,macwilliam2012scaling}.

Our paper presents a complementary approach that focuses on managing \textit{demand}.
Given that single deadlines are observed to create large spikes of \OH{} demand, splitting the deadline in half should create smaller and more manageable spikes.
We designed and implemented this policy in a large introductory computing (CS1) course at University of Illinois.
Students were split into two deadline groups, where release and due dates were staggered by one day. All students had the same amount of time to complete each assignment, but only half shared due dates.

We refer to this policy as \textit{split deadlines}.
We anticipated that it would improve our ability to support students around deadlines.
But we were concerned about fairness---both actual and perceived---and about overall student perception of the policy.
In this study, we evaluate split deadlines in three ways:

\begin{enumerate}
  \setlength\itemsep{0em}
  \item[(RQ1)] Did split deadlines improve our ability to support students in \OHS{} around deadlines?
  \item[(RQ2)] Was there a difference in student performance between the two deadline groups?
  \item[(RQ3)] How did students perceive the split deadline policy, particularly in terms of fairness and effectiveness?
\end{enumerate}

This paper makes the following contributions.
We found that split deadlines reduced \OH{} demand spikes and increased \OH{} efficiency near deadlines.
We observed no statistically significant performance differences between deadline groups.
Overall, students believed the policy was fair and effective in reducing \OH{} wait time.
We also discuss how to improve split deadline policies, based on student feedback and our own experiences.

\section{Background and Motivation}

CS1 students typically have access to a variety of support resources---including peers, websites, and \OHS{}~\cite{fong2023academic,wirtz2018resource}.
However, tutoring has been long considered effective at supporting learning, making effective \OHS{} important for CS1 despite the variety of other available support resources~\cite{wood1976role,chi2001learning}.
Doebling and Zakerouni reported that formal support like \OHS{} had a stronger correlation with undergraduate computing student success than informal support like peers and websites~\cite{doebling2021patterns}.
Gao et al. found that female students, students with fixed mindsets about computing, and students with weaker knowledge-transferring skills attended \OHS{} more frequently~\cite{gao2022uses}.
They also found that attending \OHS{} was associated with increased interest in computer science and the development of a growth mindset~\cite{gao2022uses}.
Wirtz et al. found that students tended to start with resources like peers, the course forum, and other online resources, and seek support from \OHS{} when those resources failed~\cite{wirtz2018resource}.
Overall, multiple previous studies support the role of effective \OHS{} in supporting student success in CS1.

Several previous efforts to improve \OHS{} have focused on reducing waiting and improving efficiency through the use of queue management systems.
My Digital Hand is an online tool used to manage \OHS{} at multiple institutions~\cite{smith2017my}.
In addition to maintaining a request queue, My Digital Hand also records student and staff interaction and can distribute questionnaires.
Studies of My Digital Hand have found that waiting time increases near deadlines, and that waits during busy periods can exceed one hour, sometimes causing students to leave before receiving support from course staff.
Harvard's CS50 Queue also manages a request queue, ordering student requests by creation time and allowing course staff to work together to develop an optimized schedule~\cite{macwilliam2012scaling}.
However, CS50 Queue waiting times can still grow to exceed an hour as deadlines approach.
Our \Intro{} course also uses a queue management system, and data drawn from multiple semesters confirms what previous studies have shown: Students' support requests spike before deadlines, and this can lead to long waits, overcrowded \OHS{}, and a stressful environment for staff and students.

\section{Context}
\label{sec:course_structure}

\Intro{} is an introductory computer science course at University of Illinois at Urbana-Champaign.
\Intro{} is the first course for computing majors, but also a popular service course.
Around 80\% of its yearly enrollment of nearly 2,000 students are non-majors.

Since Fall~2020, \Intro{} has not held synchronous lectures.
Students complete a series of daily lessons that combine text, code examples, interactive code walkthroughs, and video explanations.
Each lesson contains several programming exercises, one for credit and others for practice.
Each week students complete a proctored computer-based quiz.
Over the second half of the semester, students complete a longer multi-part Android programming project over a series of project checkpoints space one-to-two weeks apart.
Students can choose to complete the course in either Java or Kotlin.
This format has proven successful.
Over a typical year, around 80\% of students earn A-range grades and less than 4\% fail.

Several design choices made by \Intro{} to support student success helped motivate the split deadline policy.
First, \Intro{} aims to provide continuous support for students who need support or have questions, which is particularly important given the asynchronous format of the course.
\Intro{} \OHS{} are staffed 14~hours per day on weekdays and for more limited hours on weekends, with the goal of ensuring that anytime a student needs support, they can quickly connect with a course staff member.
Second, \Intro{} attempts to spread student workload evenly over the semester---via the daily lessons, weekly quizzes, and checkpoints included in the programming project.
Encouraging students to learn a bit each day and discouraging them from procrastinating has a variety of benefits, but also helps reduce spikes in \OH{} utilization, which can threaten the goal of providing low-latency support.
While in-person \OHS{} are offered occasionally, online support significantly reduces the overhead for students to ask questions, since they no longer need to trek across campus to a given location.

Note that \Intro{} student demographics vary between fall and spring semesters, as shown in Table~\ref{table-context}.
The presence of most computing majors in the fall semesters results in different patterns of \OH{} usage due to differences in student prior experience and motivation.
Therefore, in our analysis we compared Fall 2021 with Fall 2022, and Spring 2022 with Spring 2023.
Fall 2021 and Spring 2022 used single deadlines, whereas Fall 2022 and Spring 2023 used split deadlines.
The Android programming project used by \Intro{} also rotates yearly, meaning that Fall 2021 and Spring 2022 used one project, while Fall 2022 and Spring 2023 used another.
However, core project learning objectives are the same from year to year, and the projects are designed to have similar difficulty levels.

\begin{table}[H]
\small
\centering
\begin{tabular}{lrrlc}
\textbf{Semester} & \textbf{Enrollment} & \textbf{\% Majors} & \textbf{Deadlines} & \textbf{Project} \\
\hline
Fall 2021 & 1277 & 38 & Single & A \\
Spring 2022 & 594 & 2 & Single & A \\
Fall 2022 & 1231 & 32 & Split & B \\
Spring 2023 & 746 & 1 & Split & B \\
\end{tabular}
\vspace*{0.1in}
\caption{\textbf{Semester-by-Semester Comparison.}
Most computing majors enroll in \Intro{} in the fall semester.
Split deadlines were used in the 2022--2023 academic year.}
\label{table-context}
\vspace*{-0.3in}
\end{table}

\subsection{Defining Office Hour Demand and Efficiency}
\label{sec:definition}

\Intro{} uses its own system for managing online \OHS{}.
Students initiate requests through the website, and then wait on an embedded video call.
When a staff member is available, they enter the student's room and provide support.
When a student's request is resolved, they leave the room, closing the request.
Note that a request may not be resolved after one staff visit, in which case the student continues waiting for additional support.
\Intro{} staff are trained to rotate between requests proactively, particularly during busy periods, and students are aware that there is no expectation that the first staff visit will completely resolve their request.

We measure \OH{} \textit{demand} as the number of requests made over a particular time interval.
Because staff may not be able to resolve all requests during periods of high demand, we include both requests that were resolved and ones that were not, including requests where students were never visited by staff.
Students can and do request support multiple times within any given time interval, so the total request count is a better measure of demand than the total number of distinct students who were supported.

We measure \OH{} \textit{efficiency} as the ratio between support time and waiting time.
If a student spent 100~minutes waiting for support and 10~minutes working with a staff member, the resulting efficiency is 0.1.
Total \OH{} efficiency during a given time period is the ratio between aggregate support time and aggregate waiting time.
This definition of efficiency also helps control for variation in the project over the years that we studied.

\section{Split Deadlines}
\label{sec:split_deadlines}

Next we describe the implementation of split deadlines in \Intro{} and describe our goals when evaluating this policy.

\subsection{Implementation}
\label{sec:implementation}

Significant \OH{} demand spikes were observed near deadlines for the longer programming project assigned during the second half of each semester, as described in Section~\ref{sec:course_structure}.
Because students have some scheduling flexibility when completing the daily programming exercises, and because the exercises are smaller, they were not observed to cause problematic spikes in \OH{} demand.
(Students may not request support during quizzes, so they do not contribute to \OH{} demand.)
Therefore, we decided to implement split deadlines for project checkpoint deadlines only.

The split deadline policy was announced as students started on the longer programming project assigned in the second half of the semester.
The announcement outlined the goals and reasoning behind the policy, and explained how it was designed to help course staff better support all students.
Students were randomly assigned to the first or second deadline group, which we refer to as the Early group and Later group.
In our implementation they remained in the same group for all project checkpoints.
Students may choose to work on the project with a partner, and pairs were assigned to the same deadline group.
Deadlines for for both groups were announced to students and posted on the course website for reference.

At the end of each project checkpoint, submissions were due for Early students first and for Later students 24 hours later.
To equalize the amount of time that students had to complete each checkpoint, the test suites needed to begin each checkpoint were also provided to groups on a staggered basis.
As a result, all students had the same amount of time to work on each project checkpoint, regardless of their assigned group.

\subsection{Evaluation}
\label{sec:concern}

Split deadlines are intended to improve \OH{} efficiency by reducing spikes in \OH{} demand near assignment deadlines.
To evaluate split deadlines, we first considered effectiveness.
Did they successfully spread \OH{} demand around deadlines and reduce load spikes?
And did this lead to an increase in \OH{} efficiency, allowing students to receive more support and wait less?

We also anticipated several ways in which the policy could result in unfairness---either real or perceived.
%
%
Perhaps students with later deadlines are able to learn enough about the assignment from students with earlier deadlines to get started.
As another potential source of unfairness, note that both Early and Later group students may be receiving support through \OHS{} on the Early group deadline, which may cause longer wait times for Early group students.
On the Later group deadline, the Early group students are just beginning the next checkpoint, and are unlikely to produce significant \OH{} demand.

To fully evaluate split deadlines, we also needed to consider fairness.
Did one group end up with an actual advantage over the other, measured by their performance on the project?
And did students perceive the policy to be fair, since their perception of fairness might affect their satisfaction and persistence~\cite{chambers2023impact, wendorf2005influence}?

\section{Methods}

To understand the fairness and effectiveness of the split deadlines policy, we conducted our analysis from three perspectives.
We measured effectiveness by analyzing office hour log data and comparing \OH{} demand and efficiency between semesters with and without split deadlines.
Then we investigated students' programming assignment grades to understand whether the split deadlines policy was associated with any statistically significant differences in grades.
We also surveyed students in Spring 2023 and asked their perception of fairness and effectiveness about this policy.
In this section, we report our methods of investigation and evaluation.

\subsection{Office Hour Demand and Efficiency}

\Intro{} has its own online office hour management system, so activity logs of students and course staff were recorded by the server and used for our analysis. We analyzed office hour log data from two semesters before split deadlines (Fall 2021 and Spring 2022) and two semesters with split deadlines (Fall 2022 and Spring 2023).
We calculated \OH{} demand and efficiency (as defined in Section~\ref{sec:definition}) on each assignment due date and took the average to calculate the overall demand and efficiency metrics for each semester.
Then we compared \OH{} demand and efficiency on assignment due dates for Fall 2021 versus Fall 2022, and Spring 2022 versus Spring 2023 respectively.



\subsection{Programming Assignment Grades}


When the split deadlines policy was designed and implemented, one possible concern was whether it might provide uneven benefit or lead to disadvantages for one group over the other, which could be eventually reflected in students' programming assignment grades. As discussed in Section~\ref{sec:concern}, we suspected there might be advantages for the Later group that could lead to uneven performance on these programming assignments. Therefore, we analyzed students' programming assignment grades with statistical tests and investigated whether there was any statistically significant difference. 

To investigate potential differences in performance, we compared the programming assignment grades of students with Early deadlines and students with Later deadlines across two semesters. There were three programming assignments in both Fall 2022 and Spring 2023, and we used the average of all three assignments as our measurement. We used two-sample $t$-tests to examine whether the programming assignments grades were significantly different between Early and Later groups within each semester respectively. 


\subsection{Perception of Fairness and Effectiveness}


As explained in Section~\ref{sec:concern}, fairness was one of the biggest concerns when we designed and implemented the split deadlines policy. 
At the end of the Spring 2023 semester, we sent out an anonymous, optional survey to the class soliciting their opinions about the split deadline policy, including a five-point Likert-scale question asking students to rate their perceived fairness (``Do you think that CS124's split deadline policy was fair or unfair?''), another five-point Likert-scale question for perceived effectiveness of the policy (``Do you believe that CS124's split deadline policy reduced the wait time you experienced when seeking help on the [programming assignments]?''), and an open-response question for more insights (``What are the reasons you think the split deadline policy was fair or unfair?''). Students who filled out this survey were entered in a raffle for 20 \$10 Amazon gift cards. We collected 141 survey responses (out of 746 students). Among respondents, 68 were on the Early group and 73 were on the Later group.

We conducted Mann-Whitney $U$-tests on each of these two Likert-scale questions to examine whether there were statistically significant differences between Early and Later groups regarding perceived fairness and effectiveness. We also calculated Pearson's correlation coefficient between these two questions.

For the open-response question soliciting reasons for perceived fairness, the first and last authors conducted a thematic coding. Since this question was optional on the survey, only 110 out of 141 students answered this question. Both researchers first processed all responses independently and designed their own open-coding protocols ~\cite{corbin_basics_2008}. Then, they discussed their results and collectively compiled a code book and re-labeled all responses together, and they made sure they reached consensus on every code for every response. After the coding process was complete, the first author synthesized salient higher-level themes out of these codes.




\section{Results}
In this section, we present the results of our methodology grouped by RQs and discuss how these findings answer our RQs.

\subsection{RQ1: Did Split Deadlines Improve Our Ability to Support Students?}

To understand office hour demand, we first calculated average number of requests for online office hour help on due dates for each semester. For Fall 2021, the average number of requests was 875; whereas the average number of requests for Fall 2022 was 705 on Early deadlines and 553 on Later deadlines. Similarly, for Spring 2022, the average number of requests on due dates was 717, and the number for Spring 2023 was 343 for Early deadlines and 469 for Later deadlines. 
It appears that the absolute demand on a single deadline day decreased with the split deadlines policy, even as course enrollment increased from Spring 2022 to Spring 2023.

The average wait time per student on assignment due dates stayed roughly the same between fall semesters (57.5 minutes on Fall 2021 deadlines, 55.7 minutes on Fall 2022 Early deadlines, and 50.7 minutes on Fall 2022 Later deadlines) and decreased for spring semesters (39.1 minutes on Spring 2022 deadlines, 23.2 minutes on Spring 2023 Early deadlines, and 24.7 minutes on Spring Later deadlines). Besides, the time students spent being helped by TAs increased for both fall and spring semesters (5.3 minutes on Fall 2021 deadlines, 10.0 minutes on Fall 2022 Early deadlines and 11.1 minutes on Fall 2022 Later deadlines; 13.3 minutes on Spring 2022 deadlines, 17.8 minutes on Spring 2023 Early deadlines, and 19.3 minutes on Spring 2023 Later deadlines). 
The help-to-wait time ratio, our metric for efficiency, increased in the semesters with the split deadlines. The ratio was 0.15 on Fall 2021 deadlines, and increased to 0.35 on Fall 2022 Early deadlines and 0.41 on Fall 2022 Later deadlines. Similarly, the ratio increased from 0.70 on Spring 2022 deadlines to 1.22 on Spring 2023 Early deadlines and 1.42 on Spring 2023 Later deadlines. \textbf{We conclude that \OH{} efficiency roughly doubled on assignment due dates when the split deadline policy was in place.}

\begin{table}[H]
\begin{tabular}
{p{0.155\textwidth}|p{0.13\textwidth}|p{0.13\textwidth}}
    \toprule
    Semester (Deadlines) & Average number of help requests per deadline day & Average help-to-wait time ratio on due dates\\
    \hline
    Fall 2021 & 875& 0.15\\
    \hline
    Fall 2022 Early & 705 & 0.35\\
    \hline
    Fall 2022 Later & 553 & 0.41 \\
    \hline
    \hline
    Spring 2022 & 717 & 0.70\\
    \hline
    Spring 2023 Early & 343 & 1.22\\
    \hline
    Spring 2023 Later & 469 & 1.42\\
    \bottomrule
\end{tabular}
\vspace*{0.1in}
\caption{Metrics of office hour demands and efficiency on due dates for four semesters.}
\label{tab:oh_metric}
\vspace*{-0.2in}
\end{table}

\subsection{RQ2: Was There a Difference in Student Performance Between the Two Groups?}

\begin{figure}[H]
    \centering
    \includegraphics[width=0.45\textwidth]{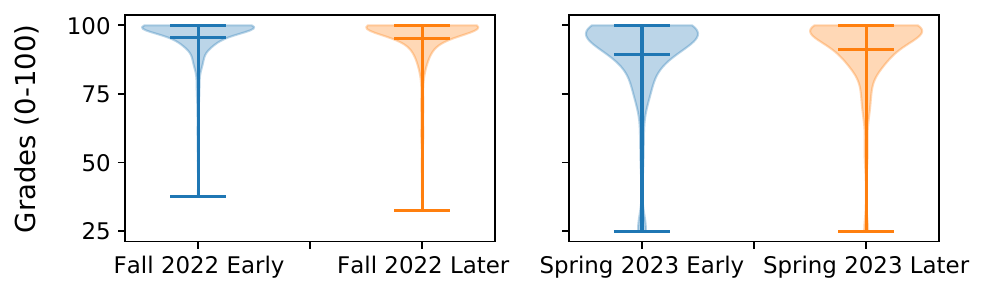}
    \caption{Students' programming assignment grades. 
    There was no statistically significant difference between Early and Later groups for both Fall 2022 and Spring 2023. Horizontal bars represent maximum, mean, and minimum scores.}
    \label{fig:mp_grades}
\end{figure}

Plots of grade distributions can be seen in Figure~\ref{fig:mp_grades}. The majority of students (85.3\% for Fall 2022, 74.6\% for Spring 2023) had an average above 90 points, and we observed ceiling effects for both semesters. For Fall 2022, the average programming assignments grades of Early group was 93.14 and 93.15 for Later, and two-sample $t$-test results ($p=0.99$) suggested there was no difference between two groups. For Spring 2023, the average was 89.25 for Early and 91.31 for Later, and $p=0.067$ from a two-sample $t$-test showed marginally non-significant difference. We also conducted two-sample $t$-tests for each individual programming assignment, and we observed no statistically significant difference for any of them. Therefore, we conclude that there was no significant difference between Early and Later groups with respect to programming assignments grades, and that the split deadlines policy was fair for students in this regard.

\subsection{RQ3: How Did Students Perceive the Split Deadlines Policy?}

\subsubsection{Quantitative analysis}
\begin{figure}[H]
    \centering
    \includegraphics[width=0.45\textwidth]{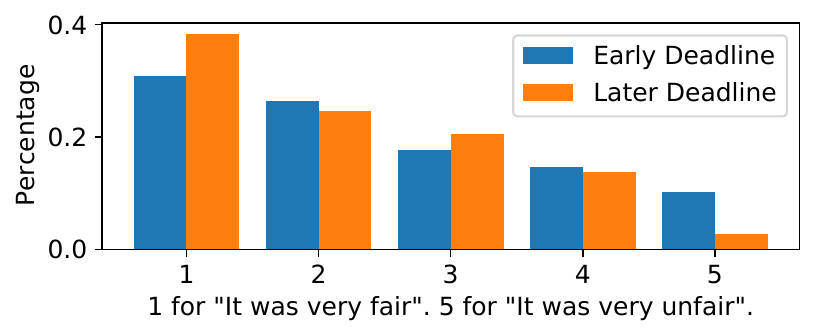}
    \caption{Students' responses to the question ``Do you think that CS124's split deadline policy was fair or unfair?''. 
    }
    \label{fig:likert-fairness}
\end{figure}

For the question ``Do you think that CS124's split deadline policy was fair or unfair?'', the Early group had a mean rating of 2.47 and standard deviation of 1.33, and the Later group had a mean rating of 2.18 and standard deviation of 1.16 (1 was  ``very fair'', and 5 was ``very unfair''). We conducted a Mann-Whitney $U$-test and there was no statistically significant difference between these two distributions ($p$=0.23). On the two ends, we observed that a higher percentage of students from the Early group believed this policy was ``very unfair'' than those from the Later group (10\% versus 3\%), and that more students from the Later group thought ``it was very fair'' than those from the Early group (38\% versus 31\%). The full distribution of responses can be seen in Figure~\ref{fig:likert-fairness}. We conclude that students believed the policy was fair, and that there was no statistically significant difference between Early and Later groups in this regard.

\begin{figure}[H]
    \centering
    \includegraphics[width=0.45\textwidth]{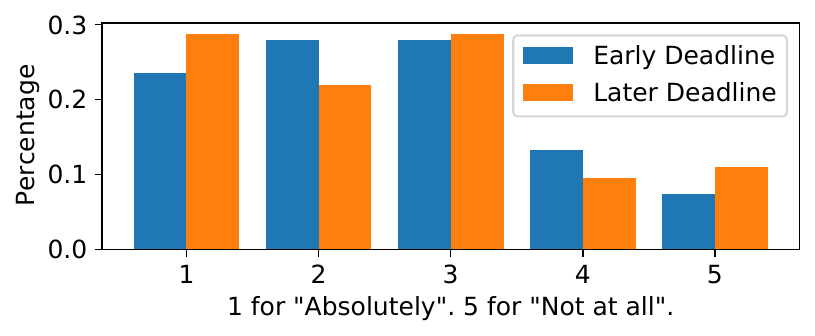}
    \caption{Students' responses to the question ``Do you believe that CS124's split deadline policy reduced the wait time you experienced when seeking help for the [programming assignments]?''. 
    }
    \label{fig:likert-effectiveness}
\end{figure}

For the question ``Do you believe that CS124's split deadline policy reduced the wait time you experienced when seeking help on the [programming assignments]?'', the Early group had a mean rating of 2.53 and standard deviation of 1.19, and the Later group had a mean rating of 2.52 and standard deviation of 1.29 (1 was  ``absolutely'', and 5 was ``not at all''). A Mann-Whitney $U$-test yielded $p=0.87$, suggesting no significant difference in perceived effectiveness between Early and Later groups. The full distribution of responses can be seen in Figure~\ref{fig:likert-effectiveness}. We conclude that students \textit{believed} that the policy was effective in reducing office hour wait time, and there was no statistically significant difference between Early and Later groups on this question.

The correlation coefficient between these two questions was 0.46, suggesting a medium-to-large positive correlation between students' perception of fairness and that of effectiveness ~\cite{cohen2009pearson}.

\subsubsection{Qualitative analysis}
The first and last author compiled a codebook with 26 codes that described the sentiments of student responses. Below are the most salient themes synthesized from the codes about why students thought the policy was fair or unfair.

\textbf{Theme 1: Procedural fairness 
was critical.
} 
Students highlighted several aspects of the implementation of the split deadlines policy as reasons for the policy's fairness. These reasons are examples of \textit{procedural fairness}, a form of justice embodied in the execution process of a policy ~\cite{van1998we}. 

One of the most popular sentiments expressed by students when describing their perception of fairness was that all students had the same amount of time (24.5\% referenced this). Students pointed out that since both groups had an identical duration to work on programming assignments, they believed split deadlines were fair. 
\begin{quote}
    I have no reasons to think its unfair. I didn't see anyone take advantage of it, and honestly, I don't know how you could. Whichever group had the later deadline also had a later start date, so its not like one group was given more time to complete the assignment, they were simply on different timelines. (P42)
\end{quote}
Students remarked that the staggered release of the test cases were key to the policy being fair (8.2\% specifically mentioned test cases). 
\begin{quote}
    It was fair because people who received the test suites one day later had an extra day for the deadline. (P76)
\end{quote}
While the equality in time 
was the most commonly expressed reason for procedural fairness, two other actions by the instructor were mentioned by students as reasons for the fairness of split deadlines. One reason was clear and proactive communication about the policy (\textit{``It was communicated prior[...],''} said P140). The other was the random assignment of students to deadlines (\textit{``It was a random split and there was no outside deciding factor,''} said P48).

\textbf{Theme 2: Effectiveness correlates with perception of fairness.}
Many students approached the question of why they believed the split deadlines policy was fair from the perspective of how it affected \OH{} waiting time. 25.5\% of students described their belief that the \OHS{} experience was improved as a result of the split deadlines policy. 
Students believed that since this policy was effective in reducing \OH{} wait time and improving their access to help during high-demand times, the policy was fair. 
\begin{quote}
The alternative scenario of having a backed-up help site is certainly less preferable to the split deadline policy. Overall, I'm glad that I was able to get help when I needed it on  [programming assignments]. (P9) 
\end{quote}
However, there were some students who believed the wait time was still long (8.2\%), so they did not believe this policy was fair. 
Students associated their perceived effectiveness of the split deadlines policy with the policy's fairness, a sentiment also communicated by the correlation between responses to the two Likert scale questions about the policy's fairness and the policy's effectiveness. 

\textbf{Theme 3: Some still believe Later is better than Early.} 
Despite the equal time students had with assignment test suites,
some students felt 
that he Early group had ``less time'' (12.7\% mentioned this). One reason for this perception was procrastination (7.3\% said).

\begin{quote}
I think it was unfair because no one really starts on the earlier day when assignments was released. Everyone pushes it till the last day. It seems like we have a day less even though we technically have the same time. But overall I think this fault is due to the students. So it is fair, but it doesn't feel fair because everyone procrastinates till the last day. (P107)
\end{quote}

Students mentioned some other reasons the Later group had an advantage, including 
that those in the Early group could 
share files or guidance with students in the Later group before they had official access (8.2\% mentioned this).

\begin{quote}
It could be considered *slightly* unfair because people at a later deadline could learn through friends or in other ways, what the goals of the project were, giving them an additional day to work on the project. (P124)

\end{quote}

A few respondents stated that as deadlines approached, the course forum became more established with questions already asked and answered and course staff more prepared,
giving some advantage to the Later group 
(1.8\% said this). 
Another issue raised (by 2.7\% of students) was that on Early deadline dates, Later students also used \OHS{} because their deadlines were upcoming soon; whereas on Later deadline dates, Early students did not have high demands for \OHS{} since they just started a new assignment. Besides, Early students also commented that they had more stress (2.7\% shared this).




\textbf{Theme 4: Splitting by only one day was important.}
Based on students’ responses, it turned out students agreed with the instructor's decision to split deadlines by \textit{only} one day (6.4\% mentioned this). When talking about the one-day difference, many students explicitly used the word ``only’’, and believed it was reason for fairness. For example,  P40 said \textit{``the idea of split deadlines was good in my opinion since it was only a difference of one day.’’} 
Students also shared their prediction of what it would be like if the split deadlines were farther apart:  \textit{``[…] it would have been in my eyes unfair if the difference was something like a week,’’} said P36.

\textbf{Theme 5: Interference with other deadlines.}
Since only programming assignment deadlines were split while other deadlines (for daily short homework assignments and quizzes) were the same for all students, split deadlines sometimes interacted differently with these other deadlines for the two groups. Some of our respondents mentioned this as a reason for unfairness (10.0\%). 

\begin{quote}
I had a class quiz class homework and [prog. assignment] due in one day whereas the other group did not have to deal with 3 major deadlines in one day.~(P39)
\end{quote} 


\textbf{Theme 6: For some, their experience didn't change.}
Some students shared that their personal experience in the class was not affected by the split deadlines policy. 8.2\% expressed this sentiment without further explanation.
Others mentioned that since there was enough time for each programming assignment (8.2\%), or that they tended to start early (1.8\%), they did not experience crowded \OHS{} and could always complete assignments smoothly.

\section{Discussion}
\label{sec:discussion}

We consider the split deadlines policy successful in \Intro{} based on our evaluation of office hour demand and efficiency and student performance and perception. In this section, we share our recommendations and reflect on some limitations.

\subsection{Recommendations}

Based on our experience and feedback from students, we provide the following recommendations for instructors considering split deadlines for their own courses. 

\begin{enumerate}
    \item \textbf{Ensure that students have equal time to complete assignments.}
    In \Intro{}, the staggered release for assignments gave each deadline group the same duration for assignments, and was a top reason students cited for fairness perceived.
    %
    \item \textbf{Explain the benefits.}
    We found that student belief that \OH{} wait times were reduced was associated with their perception that the policy was fair.
    Sharing the reasoning behind the policy---or better yet, evidence that it is working---may result in more positive views of the approach.
    \item \textbf{Carefully schedule split deadlines}.
    While separating split deadlines by more than one day may decrease demands on \OH{} staff, our findings suggest that students may find it less fair.
    Also consider how the split deadlines overlap with other course activities to avoid disadvantaging one group.
    \item \textbf{Encourage students to start early.}
    Take actions to incentive students to start early and avoid procrastination, as procrastination could undermine the effectiveness and perceived fairness of split deadlines.
    \item \textbf{Consider rotating deadline groups.}
    Rotating groups between earlier and later deadlines may help address concerns about fairness, although it may also produce confusion.
\end{enumerate}

\subsection{Limitations}

There are some threats to the validity of our results.
We did not begin studying the split deadline policy until the Spring 2023 semester, at which point it was too late to survey Fall 2022 students.
And while we received survey responses from 141 Spring 2023 students, the response rate was only 18.9\%, and so the survey might not accurately reflect the opinions of the entire class.

\section{Conclusion}

Our analysis of the design and implementation of split deadlines in \Intro{} found that the policy met its goals.
It reduced spikes in \OH{} demand and increased \OH{} efficiency around deadlines, while not causing significant performance differences or perception problems.
The majority of students perceived the policy as fair, while expressing some reasonable concerns, some of which may be addressed by future refinement of the policy.
Our experience suggests that split deadlines may be an effective way for courses to support students in \OHS{} near deadlines.

\begin{acks}
This work was supported by the National Science Foundation under Grant No. 1652397.
\end{acks}

\bibliographystyle{ACM-Reference-Format}
\bibliography{sample-base.bib}

\end{document}